\documentclass[11pt]{article}
\usepackage{fleqn,cospar}

\usepackage{url}
\usepackage{graphicx}
\usepackage[figuresright]{rotating}

\title{X-ray flares/bursts from GRS~1915+105 and the two component accretion flow}

\author{J. S. Yadav\address{Tata Institute of Fundamental Research, Homi Bhabha Road, Mumbai,
400 005, India}, and  A. R. Rao$^{1}$}

\begin{document}

% typeset front matter
\maketitle

\begin{abstract}

We present  results  of our  analysis of a set of RXTE/PCA observations of 
X-ray flares/bursts  with burst cycle ranging  from 30 to 1300 s 
from the Galactic X-ray transient source GRS 1915+105 during last 
four years.
These flares/bursts can be classified into four 
types: 1) regular bursts with short burst phase of $\sim$ 20 s (R1), 
2) regular bursts with short quiescent phase of $\sim$ 20 s (R2), 3) 
long regular bursts (R3) with burst cycle $\ge$ 1000 s, and 4) irregular 
bursts (IR). For all the observed bursts,   the duration of the quiescent  
phase is inversely proportional to the square of the QPO frequency (2$-$10 Hz).
\end{abstract}

\section{Introduction}
The Galactic X-ray transient source GRS~1915+105 has shown spectacular 
X-ray variability during last four years of its observation by RXTE and other 
satellites (Greiner et al., 1996, Yadav et al., 1999, Belloni et al., 2000). 
This source  was 
discovered in 1992 with the WATCH 
all sky X-ray monitor onboard the GRANAT satellite (Castro-Tirado et al. 1994).
 The X-ray intensity was found to vary on a variety of time scales and
the light curve showed a complicated pattern of dips and rapid transitions 
between high and low intensity (Belloni et al., 1997, 
Taam et al., 1997).
 Recently, Yadav
et al. (1999)  have made a detailed study of various types of X-ray bursts
seen in GRS~1915+105 from IXAE/PPCs observations during 1997 June - August  
and  attempted to explain  these bursts
in the light of the recent theories of advective accretion disks.
In X-ray astronomy, the terminology X-ray burst is  used  for type I and 
type II classical bursts seen in LMXBs. The timing and spectral properties of 
the flares/bursts described here are completely different than that of the
classical bursts (Yadav \& Rao, 2000).
\begin{table}[b]
\begin{center}
\centerline{Table I.}
\centerline{Summary of selected observations of GRS~1915+105} 
%\small
\begin{tabular}{lclccc}
\hline\hline     
 Date &   Exposure & Type of  &ASM$^a$ & Rec.&Av. Q.\\ 
  (UT) &   (s) & Bursts & Flux &Time (s)&Flux\\
\hline  
 1996 April 06 05:40& 5600 &Regular (R2)&99.4&280$^{b}$&6200\\ 
 1996 Oct 07 05:44 & 7000 &Regular (R3)&98.8&1150$^{b}$&3200\\ 
 1997 May 26 12:25 & 3215 &Regular (R1b)&47.6&105$^{b}$&7500\\ 
 1997 June 18 14:17 & 3472 &Irregular (IR)&61.5& var.& var.\\ 
 1997 June 22 19:27 & 2550 &Regular (R1a)&59.7&55$^{b}$&8700\\ 
\hline  
\end{tabular}
\end{center} 
{Q. Flux= average quiescent time flux (c/s), var. = variable}, 
$^{a}${Mean ASM flux (c/s) for a day},  
$^{b}${Mean burst recurrence time}.
\end{table}
%\section*{OBSERVATIONS}

\begin{figure}[t]
\centering
\includegraphics[width=100mm,angle=-90]{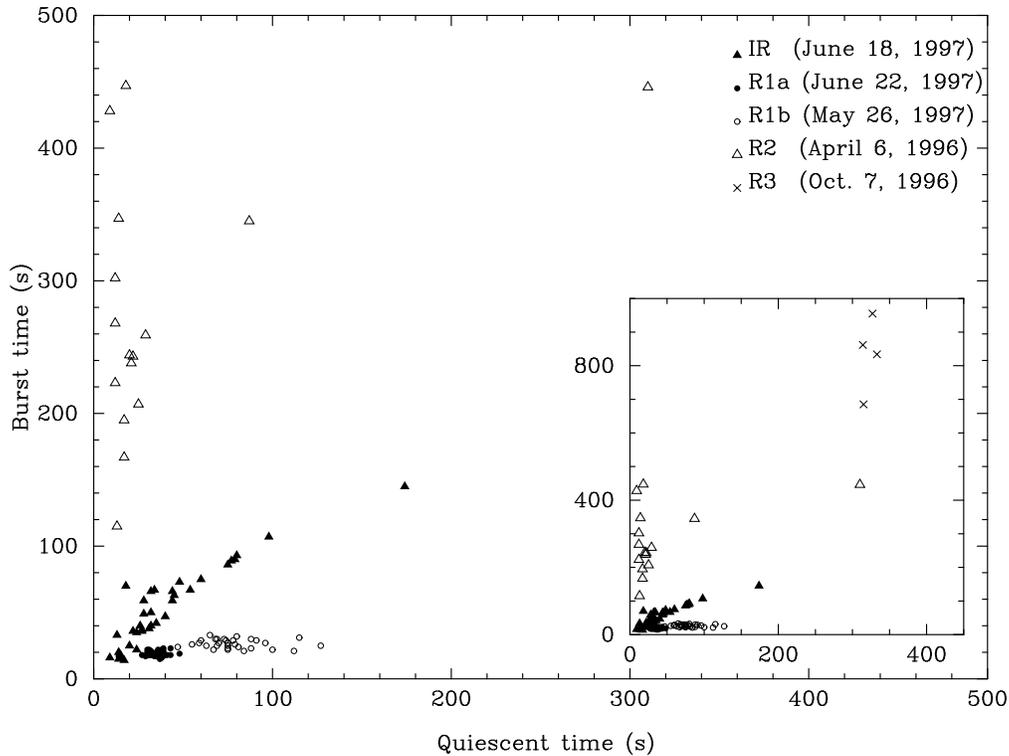}
\caption { Plot of the  quiescent time  and the burst duration 
 for all the  bursts.}
\end{figure}

 In Table 1, we list details of RXTE/PCA  observations discussed here
along with some of the
properties of the observed X-ray flares/bursts.
  Each burst cycle consists of a low flux 
quiescent phase
followed by a high flux burst phase and the fast transition in less than 10 s.
The dips or the quiescent phase in 
these observations are spectrally hard while the brighter portions (the
burst phase) are soft (Yadav, 2001). For details of these observations
as well as of our analysis please see Yadav \& Rao (2000).

\section*{Results and discussion}
 On the basis of timing properties, these flares/bursts can broadly be put into 
two classes: regular bursts centered 
around a fixed period with  low dispersion ($\delta P / P \sim 1 - 50 \%$)
and irregular bursts with no fixed periodicity ($\delta P / P >  50 \%$).
 These flares/bursts can be classified into four 
types: 1) regular bursts with short burst phase of $\sim$ 20 s (R1) with
$\delta P / P \sim 1$-$10\%$, 
2) regular bursts with short quiescent phase of $\sim$ 20 s (R2) with
$\delta P / P$  up to 50$\%$, 3) 
long regular bursts (R3) with burst cycle $\ge$ 1000 s and $\delta P / P$
 $\sim$ 1$-$15\%, and 4) irregular bursts (IR) (see figure 1). 
We have measured the quiescent time and the burst duration for all types of 
bursts  and results are shown in Figure 1.
The data of IR bursts seen on  1997 June 18 fall diagonally; clearly 
showing a strong correlation between the burst duration and the 
quiescent time. On the other hand, data for different types of regular
bursts fall on either horizontal or vertical branches implying no such
correlation for the regular bursts (Yadav et al., 1999). The R1a and R1b 
regular bursts fall on the horizontal branch with the
burst time $\sim$ 20 s. The R2 regular bursts  fall on
the vertical branch with the quiescent time $\sim$ 20 s while the R3 regular 
bursts  fall on vertical branch with  the quiescent 
time $\sim$ 320 s(this is  shown in the inset of Figure 1).

\begin{figure}[t]
\centering
\includegraphics[width=90mm,angle=-90]{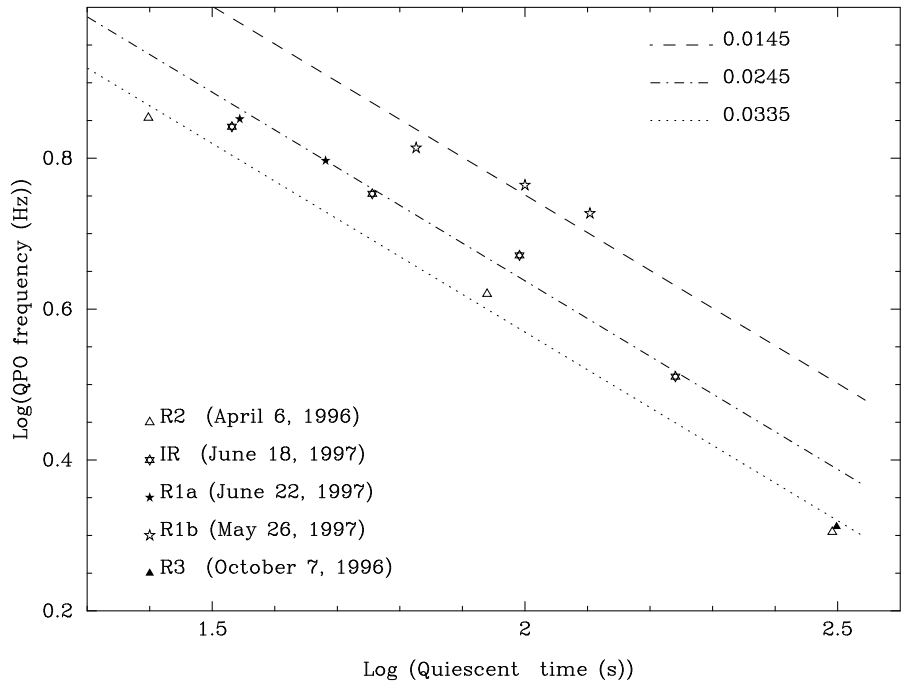}
\caption {Variation of QPO frequency $\nu_I$ (minimum) with the quiescent time for different types of X-ray flares/bursts (data
points). Plotted lines are the quiescent time $\alpha$ $\nu_{I}^{-2}$ for
different values of ${\Theta}_{\dot{M}}$.}
\end{figure}

The most striking features of these  flares/bursts  are 
slow exponential rise, sharp linear decay and hardening of spectrum as
burst progresses (Paul et al. 1998). The decay time scales are shorter than 
the rise time scales.
In sharp contrast, the decay time is longer than the rise time  in 
classical bursts and  spectrum is initially hard and becomes softer as
the burst decays (Lewin et al., 1995).
Yadav et al.  (1999)   have  suggested that the source is 
in a high-soft state during the 
burst phase and in a low-hard state during the quiescent phase  on the basis of available spectral observations and derived disk parameters of 
GRS~1915+105.
 The fast time scale for the transition of the 
state is explained by invoking the appearance and disappearance of the 
advective disk in its viscous time scale.
 Such fast changes of 
states are possible in the Two Component Accretion Flows (TCAF) where the 
advective disk covers the standard thin disk (Chakrabarti, 1996, Chakrabarti 
and Titarchuk, 1995).  
Chakrabarti \& Manickam (2000) have provided a relation; 
$t_{off}= f(\Theta_{\dot{M}})  \nu_{I}^{-2}$ based on TCAF. Where  
 t$_{off}$ is  the duration of 
off state (duration in which the sonic sphere becomes ready  for
catastrophic Compton cooling),   
  $\nu_I$ is the intermediate QPO frequency between 2 $-$
10 Hz, and $\Theta_{\dot{M}}$ is a dimensionless 
parameter defined as
$\Theta_{\dot{M}} = (\Theta_{out} / \Theta_{in}) \times \dot{m}_d$ where
$\Theta_{in}$ and $\Theta_{out}$  are the solid angles  of the inflow  \& 
outflow respectively and $\dot{m}_d$ is the disk accretion rate in units 
of Eddington accretion rate.  

 In Figure 2, we plot the above relation in the log - log scale taking t$_{off}$ 
as the quiescent time for $\Theta_{\dot{M}}$ $=$ 0.0145, 0.0245 and 0.0335
alongwith our QPO results (points).
 These results are in good agreement. 
The data points 
of R2 and R3 bursts when ASM flux was 99.4 \& 98.8 c/s respectively fall
along the dotted line with $\Theta_{\dot{M}} =$ 0.0335 (see table 1). 
The data points
of R1a and IR bursts when ASM flux was 59.7 and 61.5 c/s respectively
fall along the dashed -  dotted line ($\Theta_{\dot{M}} =$ 0.0245). The 
data points of R1b bursts  during which ASM flux has lowest value of 47.6 c/s
fall along the dashed line ($\Theta_{\dot{M}} =$ 0.0145). It may be 
noted here that the $\Theta_{\dot{M}}$ and the ASM flux though determined
independently agree well for different types of flares/bursts  as both of these
are related to the disk accretion rate $\dot{m}_d$. Our results in Figure 2 suggest that the t$_{off}$ represents the quiescent 
time of the flares/bursts which may or
may not be of the order of the viscous time scales of the thin accretion
disk.

\end{document}